\documentclass[iucr,preprint,showpacs,superscriptaddress,amsmath,amssymb,amsfonts,floatfix,longbibliography]{revtex4-1}

\usepackage{graphicx}
\usepackage{placeins}
\usepackage{float}
\usepackage{dcolumn}
\usepackage{bm}
\usepackage{amssymb}
\usepackage[version=4]{mhchem}
\usepackage{microtype}
\newcolumntype{C}[1]{>{\centering\arraybackslash}p{#1}}
\usepackage{xfrac}
\usepackage{gensymb}
\usepackage{xcolor}
\usepackage{multirow}
\usepackage{comment}
\usepackage{times}
\usepackage[varg]{txfonts}
\usepackage{mathrsfs}
\usepackage{amsmath,amssymb,bm,mathtools}
\usepackage{amsfonts}
\usepackage{booktabs}
\usepackage{natbib,hyperref}
\hypersetup{
	colorlinks=true, 
	linkcolor=blue,  
}
\usepackage{soul}
\usepackage{epstopdf}
\usepackage{array}
\usepackage{multirow}
\usepackage{tcolorbox}
\usepackage[percent]{overpic}
\usepackage{bbm}

\usepackage{widetable}
\usepackage{array}
\usepackage{xcolor,colortbl}
\usepackage{lipsum}
\usepackage{tablefootnote}

\newcommand{\ket}[1]{\ensuremath{|#1\rangle}}
\newcommand{\bra}[1]{\ensuremath{\langle#1|}}

\newcommand{\etal}{\textit{et al.}}
\newcommand{\yto}{Yb$_2$Ti$_2$O$_7$ }
\newcommand{\EMSO}{Er$_3$Mg$_2$Sb$_3$O$_{14}$ }

\begin{document}

\title{CrysFieldExplorer: a software for rapid optimization of crystal field Hamiltonian}

\author{Qianli Ma}
\affiliation{Neutron Scattering Division, Oak Ridge National Laboratory, Oak Ridge, Tennessee, USA,37831}

\author{Xiaojian Bai}
\affiliation{Neutron Scattering Division, Oak Ridge National Laboratory, Oak Ridge, Tennessee, USA,37831}
\affiliation{Department of Physics and Astronomy, Louisiana State University , Baton Rouge, Louisiana, USA,70803}

\author{Erxi Feng}
\affiliation{Neutron Scattering Division, Oak Ridge National Laboratory, Oak Ridge, Tennessee, USA,37831}

\author{Guannan Zhang}
\affiliation{Computer Science and Mathematics Division, Oak Ridge National Laboratory, Oak Ridge, Tennessee, USA,37831}

\author{Huibo Cao*}
\affiliation{Neutron Scattering Division, Oak Ridge National Laboratory, Oak Ridge, Tennessee, USA,37831}

\date{\today}


\begin{abstract}
We present a new lite python-based program, CrysFieldExplorer, for fast optimizing crystal electric field (CEF) parameters to fit experimental data. The main novelty of CrysFieldExplorer is the development of a unique loss function, referred to as the Spectrum-Characteristic Loss ($L_{\text{Spectrum}}$), which is defined 
based on the characteristic polynomial of the Hamiltonian matrix. Particle Swarm Optimization and Covariance matrix adaptation evolution strategy are used to find the minimum of the total loss function. We demonstrate that CrysFieldExplorer can performs direct fitting of CEF parameters to any experimental data such as neutron spectrum, susceptibility, magnetizations etc. CrysFieldExplorer can handle a large amount of none-zero CEF parameters and reveal multiple local and global minimum solutions. Detailed crystal field theory, description of the loss function, implementation and limit of the program are discussed within context of two examples.

\end{abstract}

\maketitle

\section{Introduction}

Single-ion magnetic anisotropy is one of the key elements for exotic quantum states in frustrated systems \cite{PhysRevB.62.6496,Gardener,huiboPRL,Bertin_2012,JoYb2015,zhilin,paddison2017continuous,JoPRL}. Crystal electric field (CEF) is responsible for single-ion magnetic anisotropy, and it occurs through Coulomb interaction and Pauli exclusion between the central cation and surrounding anions that splits the energy levels of electrons grouped by different orbitals of the central cation. In the absence of the CEF effect or in a spherically symmetrical field, these energy levels would otherwise be degenerate. The CEF effect is known to cause dramatic magnetic anisotropy which collectively produce exotic ground states in a wide range of quantum materials \cite{Gingras2001,kate2011,JoPRL,cefcehfo,keimer2017physics,evanprx} and unconventional high temperature superconductors \cite{fulde1970superconductors,mesot1997crystal,hightccef19809,metoki2004magnetic, CEF2004hightc,hightccef1, boothroydcef,pickett2021dawn}. 

The calculation of excited energy levels of magnetic ions in a crystalline environment has been well studied \cite{Stevens_1952, hutchings1964point,lea1962raising, walter1984treating,loewenhaupt1990crystal}. Stevens and Hutchings have illustrated systematically the process of determination of the perturbing Hamiltonian from the evaluation of the electrostatic potential experienced by the magnetic ion from the surrounding charges \cite{Stevens_1952,hutchings1964point}. One of the common conventions to express the CEF Hamiltonian is using Stevens Operators. The crux of the CEF study is to solve the single-ion Hamiltonian by matrix diagonalization and retrieve eigenvalues and eigenfunctions from the CEF Hamiltonian to fit the corresponding crystal field excitations. Inelastic neutron scattering is a suitable experimental technique because it directly measures transitions between ground states and excited states of magnetic ions. It has gradually become one of the most popular experimental methods in conducting crystal field analysis. After obtaining neutron scattering and other bulk property data, the next step in the CEF analysis is fitting the CEF Hamiltonian to experimental observables. It has always been a challenging optimization problem to find a solution set that best describes the experimentally measured data. Efforts to tackle this problem include
programs such as SPECTRE \cite{spectre}, McPhase \cite{McPhase}, Mantid \cite{Mantid}, SIMPRE \cite{baldovi2013simpre} CFca \cite{CFcal}, FOCUS \cite{fabi1995focus} and the latest PyCrystalField \cite{scheie2021pycrystalfield}. These programs use various techniques to calculate the CEF parameters but they all share the same approach that uses $\chi^2$, a type of loss function, to find the global minimum. One of the disadvantages associated with choosing the $\chi^2$ loss function is it relies on a set of starting parameters that needs to be adjacent to the true solution. As we will demonstrate in the next chapter, large energy boundaries in the $\chi^2$ type loss function exist that could trap the optimization process on a local minimum. Typically, the point-charge model and Monte Carlo sampling are used to generate starting parameters. However, the point-charge model is a classical approximation by positioning several point charges around a magnetic ion to represent a distribution of valence electrons. This heavily limits the accuracy of the point-charge model. Additionally, building a point-charge model or Monte Carlo simulation can sometimes be none-trivial for inexperienced researchers. More importantly, when dealing with low local symmetry such as C$_i$, C$_1$, the number of none-zero crystal field parameters can be well into high 20s. The complexity and cost of computation of the multi-dimensional loss function increases exponentially and can easily overwhelm currently existing software. 

To tackle these challenges, we have designed a special loss function, Spectrum - Characteristic Loss ($L_{\text{Spectrum}}$), based on the theory of polynomial characteristics and developed a lite python program called CrysFieldExplorer. CrysFieldExplorer takes advantage of particle swarm optimization (PSO) \cite{PSO} and covariance matrix adaptation evolution strategy (CMA-ES) \cite{hansen2006cma} to minimize a combination of Spectrum - Characteristic Loss and traditional $\chi^2$ losses. It is able to quickly fit neutron spectroscopy data and any other experimental results such as susceptibility, magnetization, specific heat, neutron diffraction, etc with CEF model and yields a series of solutions containing information with local and global minima. It bypasses the step of using a point-charge model to estimate starting parameters. In the following chapters, we will show our methods can not only discover the solutions reported in literature, but also uncover multiple other solutions that could fit all provided experimental data equally well. This suggests that the data used in the traditional optimization process may not have enough resolution to distinguish one from another and require further measurements such as local magnetic susceptibility with polarized neutron diffraction \cite{huiboPRL}. Our findings suggest that the traditional standard of determining one set of best CEF parameters may be flawed. We, therefore, suggest that future CEF analysis should include a list of possible solutions and discuss their relative physical meaning and the reasons for the preferred solution.

\section{Theory}

\subsection{Crystal Electric Field Theory}
The CEF potential can be constructed from a point-charge perspective detailed in Hutching's work \cite{hutchings1964point}. The electrostatic potential due to the surrounding point-charges can be expressed using tesseral harmonic functions in Cartesian coordinates. Following this convention, the Crystal Field (CF) Hamiltonian can be expressed by Stevens operators as \cite{Stevens_1952}:
\begin{equation}
    H_{CF} = \sum_{n,m} B^{m}_{n}O^{m}_{n}=\sum_{n,m}[A^m_n \theta_n]O^{m}_{n},
    \label{hamiltonian}
\end{equation}

where $B^{m}_{n} (|m| \leq n)$ is the CF parameters fitted from experimental measurements, $O^{m}_{n}$ are the Stevens' Operators. The CF parameters can also be expressed in terms of $A^m_n$ and $\theta_n$, where $\theta_n$ represents reduced matrix elements and is also called as the Stevens factor. For rare earth ions, $\theta_n$ has been tabulated in Table VI of Ref. \cite{hutchings1964point}.

\subsection{Neutron Scattering Cross-section}
Inelastic neutron scattering spectroscopy is well-suited in studying the crystal field excitations because neutrons can directly excite electron spins from one level to another and measure the difference between two energy levels.  The observed intensity and excited energy levels can be used to fit the crystal field parameter $B^m_n$s in Eq. \ref{hamiltonian}. The excited energy levels correspond to the energy difference between the lowest eigenvalues and the corresponding state of the CEF Hamiltonian. The observed intensity is related to the partial differential magnetic cross-section expressed as:
\begin{equation}
    \frac{d^2\sigma}{d\Omega dE'} = C\dfrac{k_f}{k_i} F(|Q|)S(|Q|,\hbar\omega),
    \label{partial}
\end{equation}
where $k_f$ and $k_i$ are the momentum of scattered and incident neutrons, C is a constant including the Debye-Waller factor, $F(|Q|)$ is the magnetic form factor of the sample. $S(|Q|,\hbar \omega)$ is the scattering function, $\hbar\omega$ indicate the energy of neutrons.  From $S(|Q|,\hbar \omega)$ the relative intensities from different states of the CEF exictations can be calculated. At constant $|Q|$, the $S(|Q|, \hbar\omega)$ is,
\begin{equation}
    S(Q, \hbar \omega)=\sum_{i,i'} \dfrac{(\sum_{\alpha}|\bra{i}J_{\alpha}\ket{i'}|^2)e^{-\beta E_i}}{\sum_j e^{-\beta E_j}} L(\Delta E+\hbar\omega), 
    \label{Sq}
\end{equation}
$i \rightarrow i'$ indicates the transition between state $i$ and $i'$ for the magnetic ion. $\beta=-\dfrac{1}{k_B T}$, $k_B$ is the Boltzman constant, T indicates temperature. $L(\Delta E+\hbar\omega)$ is a Lorentzian function that guarantees the energy conservation when neutrons induce a transition from state $i$ to $i'$, which poses a finite energy width or lifetime. The $\sum_\alpha$ sums all three $x,y,z$ component of the $J_\alpha$ matrix. Although many CEF experiments are measured at low temperature, high temperature measurements can be important for confirming or extracting the scattering signal originated from CEF excitations that can be calculated with Eq. \ref{Sq}.

\subsection{Magnetization and Susceptibility}

Magnetization under an external magnetic field $\mathbf{H}$ can be readily calculated by combining CEF Hamiltonian and the Zeeman term. The overall Hamiltonian can be expressed as:
\begin{equation}
H=H_{CF} - \mu_B g_J \bold{H}\cdot\bold{J}.
\label{Hequation}
\end{equation}
By diagonalizing Eq.\ref{Hequation}, we can calculate the eigenstates $E_n$ and eigenfunctions $\ket{i}$ of the CF Hamiltonian in a magnetic field $\mathbf{H}$. The three components ($\alpha = x,\ y,\ z$) of the magnetization in Cartesian coordinate system are given as,
\begin{equation}
    M_{\alpha}(\mathbf{H},T)=g_j\sum_n e^{-\beta E_n}\bra{n}J\ket{n}/Z
    \label{magnetization}
\end{equation}
where $Z=\sum_n e^{-\beta {E_n}}$ is the partition function. The powder average of the magnetization can be derived from Eq. \ref{magnetization} by calculating the averaged magnetization on a unit sphere. 

The magnetic susceptibility can be calculated by taking derivatives of the magnetization $M_{\alpha}$,

\begin{equation}
    \chi_{\alpha\beta}=\dfrac{\partial M_{\alpha}}{\partial H_{\beta}}.
\label{chi}
\end{equation}
The powder averaged magnetic susceptibility can be calculated in the same fashion by substituting $M_{\alpha}$ with its powder-average version. Detailed information about the calculation can be found in ref.\cite{zhilin}.

\subsection{Loss Function}

\begin{figure}
    \centering
    \includegraphics[width=6.5in]{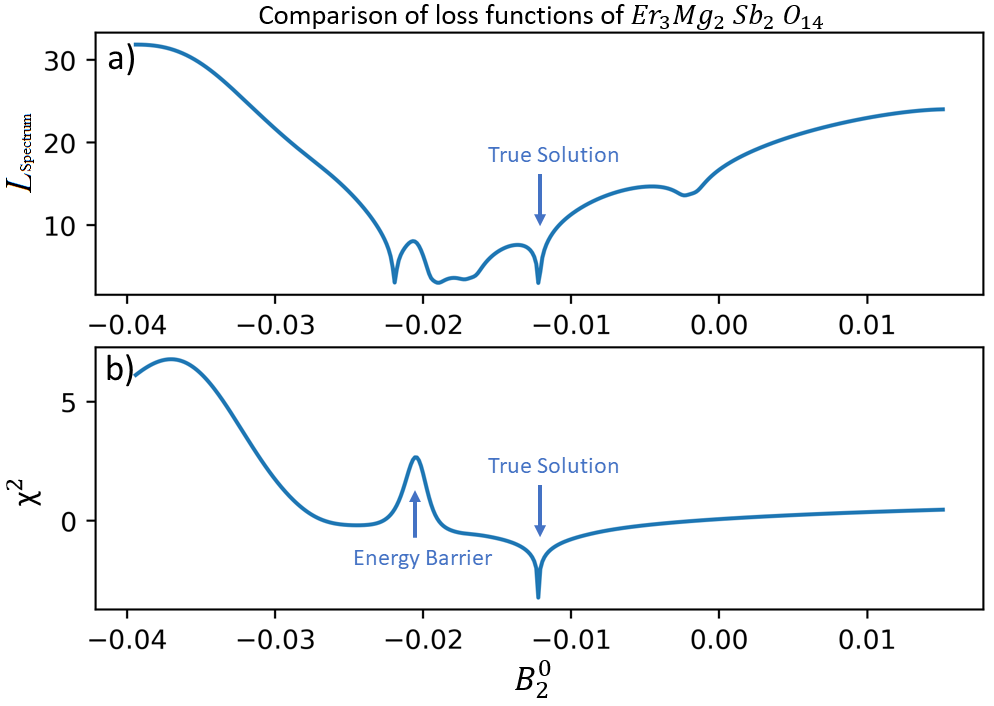}
    \caption{\EMSO comparison of Spectrum - Characteristic Loss and the mean square root loss functions commonly used in other optimization programs. Both losses are calculated along the line in a 15-dimensional parameter space and projected on the B$^0_2$ dimension.}
    \label{Loss}
\end{figure}

The core of an optimization problem is constructing a proper loss function that is smooth on the parameter space, sensitive to the change of input parameters and does not produce large energy boundaries around local and global minima. Currently, a few popular softwares such as SPECTRE \cite{spectre}, PyCrystalField \cite{scheie2021pycrystalfield} and Mantid \cite{Mantid}, etc are able to fit the CEF Hamiltonian using experimental observables. Different to the existing software packages, here we introduce a newly designed loss function, Spectrum - Characteristic Loss. and demonstrate its advantage when deployed in CEF optimization.
  
The foundation of the Spectrum - Characteristic Loss is from the theory of characteristic polynomial - $det(\lambda I - A) = 0$, where $\lambda$ is the eigenvalue of matrix A, $I$ is an identity matrix. The difference between the lowest and excited eigenvalues are neutron observed energy levels. Using this feature, we construct the loss function for the neutron measured energy levels as:
\begin{equation}
    L_{E} = \log_{10}\left(\sum_{i=1}\dfrac{det\{(E_{\text{exp}}[i]+E_{\text{cal}}[0])\mathrm{I}                 -H \}^2}{det\{E_{\text{exp}}[i]I \}^2}\right),
    \label{lossE}
\end{equation}

$E_{\text{exp}}$ indicate the observed excitation levels from neutron scattering, the summation of $i$ starts from 1 which is the first excited energy level observed by neutron experiment. $E_{\text{cal}}[0]$ is the ground state eigenvalue from matrix $H$. $E_{\text{cal}}[0]$ may not be 0.

Subsequently, we construct the mean square root deviation for the intensity and the inverse susceptibility as:
\begin{equation}
    L_{\text{Intensity}} = \dfrac{\sqrt{\sum({I_{\text{true}}[i]-I_{\text{calc}}}[i])^2}}{\sqrt{\sum{(I_{\text{true}}}[i])^2}},
    \label{lossI}
\end{equation}
and
\begin{equation}
    L_{1/\chi} = \dfrac{\sqrt{\sum({1/\chi_{\text{true}}[i]-1/\chi_{\text{calc}}}[i])^2}}{\sqrt{\sum{(1/\chi_{\text{true}}}[i])^2}}.
    \label{lossX}
\end{equation}

To construct the Spectrum - Characteristic loss function, we combine Eq.\ref{lossE}, \ref{lossI} such that $L_{\text{Spectrum}}$=$log_{10}(L_{E})+L_{\text{Intensity}}$. To fit other bulk measured physical properties, CrysFieldExplorer sums up $L_{\text{Spectrum}}$ with other mean square root losses computed from the experimental data and conduct global optimization.

Figure \ref{Loss} compares the Spectrum - Characteristic Loss with traditional mean square losses of the excited energy levels and relative intensities for an example material \EMSO. It represents an optimization conducted on a 15-dimensional parameter space. More physical details about the fitting results are discussed in a later chapter. Here we show the computed loss along a straight line passing through the true solution of \EMSO\cite{zhilin}. The line is chosen to point in a random direction within the 15-dimensional parameter space to simulate an optimization procedure (a real optimization may not follow a straight line). The loss is then projected onto the B$^0_2$ dimension for visualization. The upper panel shows the Spectrum - Characteristic Loss function while the lower panel is the standard $\chi^2$ loss function. To compare the loss functions like-to like we also apply a $log10$ operation on the energy levels such that $\chi^2=log_{10}(\chi^2_E)+\chi^2_I$. Comparing the upper panel to the lower panel, one can tell that the Spectrum - Characteristic Loss has a global structure around the minimum and does not produce sharp energy barrier at B$^0_2 \sim$ -0.02. Such an energy barrier can trick the optimization algorithm assuming it has already reached the global minimum. This can lead to a wrong solution if the starting parameters fall within the energy well. The Spectrum - Characteristic Loss also appears to be more sensitive to slight perturbations in CEF parameters. In the specific direction that the loss is computed, traditional $\chi^2$ loss function is only able to converge at the true solution which is artificially designed such that the energy levels exactly match between calculated and observed, causing both $L_{\text{Spectrum}}$ and $\chi^2$ to appear sharp at the true solution. On the other hand, the Spectrum - Characteristic Loss function can discover at least three other minima in the parameter space. In reality the real global minimum is unknown for a measured material, so it is important to have a complete list including all potential minima for further examination as different CEF parameters from the list could all reproduce results identical to experimentally measured properties. As the number of CEF parameters increases, it requires more constraints to conclusively solve the CEF Hamiltonian. The multiple solutions CrysFieldExplorer discovered is an indication that the input observables do not impose enough constraints to reach the true solution. 

In addition, CrysFieldExplorer can add any other experimental data such as specific heat, anisotropy g-tensors, magnetization, etc in the optimization process. The loss function for additional data can be added the same way as mean square root deviation. Users can also adjust weights for different losses to fine tune the optimization process.

\section{Examples}
\subsection{Benchmark with Rare Earth Pyrochlore: $Yb_2Ti_2O_7$}

Geometrical frustration has been of great interest in condensed matter physics due to the ability to host rich order-disorder states. These exotic states of matter are results of collective behaviors that arise from frustrated interactions within the quantum many-body system. Understanding these phenomena is crucial for developing next generation quantum technologies \cite{tokura2017emergent,ball2017quantum,giustino20212021,cava2021introduction,bassman2020domain,lau2020emergent,stanev2021artificial}. Rare earth pyrochlore is the archetype of magnetic frustration in three dimensions. This system has been reported to host exotic magnetic states such as quantum spin ice \cite{kate2011,kimura2013quantum,petit2017pr227}, quantum spin liquid \cite{gingras2014quantum,orderbydisorder,evanprx,bhardwaj2022sleuthing} etc. The pyrochlore structure generally possesses chemical composition of A$_2$B$_2$O$_7$ (B= Sn, Ge, Pt, Zr, Ti etc) with a space group $Fd\Bar{3}m$ and point group $D_{3d}$ at A site. We selected this system as a high symmetry end to benchmark our optimization model. Its local site possesses a 3-fold rotation axis along the local [111] direction. Due to the symmetry constraints, the number of none zero CEF parameters is 6. In this chapter, we demonstrate the use of CrysFieldExplorer by comparing a list of optimized results with observables in typical measurements such as inelastic neutron scattering, susceptibility, magnetizations and anisotropy g-tensors.

\begin{figure}
    \centering
    \includegraphics[width=5in]{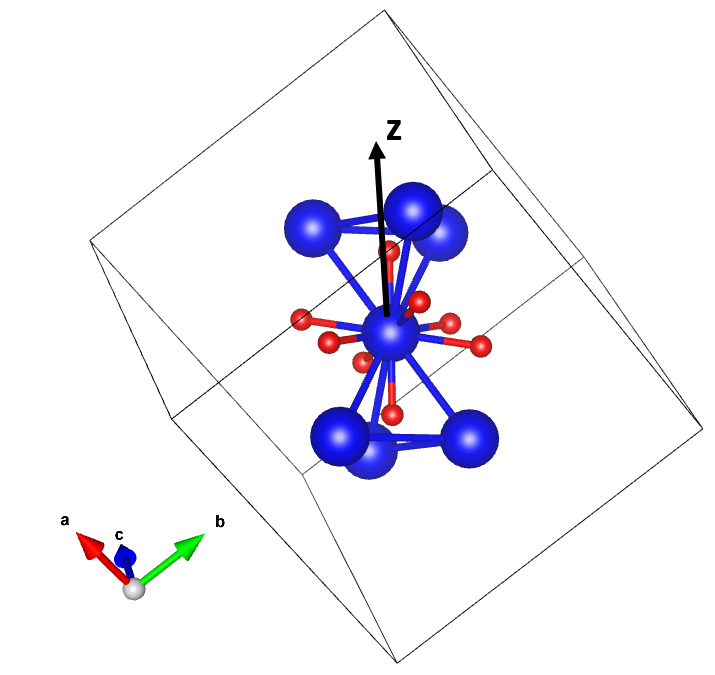}
    \caption{The corner-sharing tetrahedral structure of Yb$_2$Ti$_2$O$_7$. The blue spheres indicate Yb$^{3+}$ ions and the red spheres are the surrounding oxygen atoms around one Yb$^{3+}$ site. The z-axis is the three-fold rotation axis along the  $<$111$>$ direction.}
    \label{ybstructure}
\end{figure}

One of the representative systems is the Yb$_2$Ti$_2$O$_7$, a promising quantum spin ice candidate with XY-type single-ion magnetic anisotropy \cite{Cao_2009,huiboPRL,kate,JoYb2015}.  The ground states and crystal field effect of Yb$^{3+}$ have been well studied by Gaudet \textit{et al.} in Ref.\cite{JoYb2015}. Yb$^{3+}$ has an electronic configuration of 4$f^{13}$, total angular momentum $\mathbf{\textit{J}}=7/2$. The crystal electric field effect originates from the surrounding O$^{2-}$ ions of the Yb$^{3+}$ site. As a result, $2J+1=8$ - fold degeneracy is expected to be lifted by the CEF effect from the ground state,  resulting in 7 excited levels. Additionally, Yb$^{3+}$ is a Kramers' ion, therefore the 8 - fold degeneracy are all doubly degenerate into 4 well separated doublets. A total of 3 excitations are expected to be observed through inelastic neutron scattering spectroscopy. Oxygen atoms around the local A-site form a distorted cubic structure shown in Figure \ref{ybstructure}. The most convenient placement of the local coordination system placing the local $z$ - axis along the local three-fold $<$111$>$ rotational axis perpendicular to the oxygen plane. The resulting Hamiltonian takes the form of:

\begin{equation}
    H_{CEF}= B^0_2\hat{O}^0_2 + B^0_4\hat{O}^0_4 + B^3_4\hat{O}^3_4 + B^0_6\hat{O}^0_6 + B^3_6\hat{O}^3_6 + B^6_6\hat{O}^6_6.
    \label{Ybhamiltonian}
\end{equation}

Eq. \ref{Ybhamiltonian} follows the convention of Stevens operators for $\hat{O}^m_n$, $B^m_n$s are CF parameters used to describe the Coulomb potential generated by the surrounding oxygen atoms. Inelastic neutron scattering results have been reported in Ref.\cite{JoYb2015}. 

\setlength{\tabcolsep}{15pt}
\def\arraystretch{1.25}
\begin{table*}[]
    \centering
    \begin{tabular}{c c c c c }
    \hline\hline
        B$^m_n$ (meV) &  A.Bertin\cite{Bertin_2012} & J.Gaudet\cite{JoYb2015} & This work \footnote{The parameters indicated by green circles in Figure \ref{Ybcefparameter}} & This work \footnote{The parameters indicated by black circles in Figure \ref{Ybcefparameter}}\\
    \hline
         B$^0_2$ & 1.270 & 1.135 & 1.57 & -2.22\\
         
         B$^0_4$ & -0.0372 & -0.0615 & -0.0365 & 0.00241\\

        B$^3_4$ & 0.275 & 0.315 & -0.627 & -0.0612\\

        B$^0_6$ & 0.00025 & 0.0011 & 0.00184 & -0.00102\\

        B$^3_6$ & 0.0023 & -0.037 & -0.0285 & 0.0287\\

        B$^6_6$ & 0.0024 & 0.005 & -0.0211 & -0.0368\\

        g$_{x(y)}$ & 4.09(2) & 3.69(0.15) & 3.72 & 2.43\\
        g$_{z}$ & 2.04(3) & 1.92(0.2) & 2.07 & 4.51 \\
    \hline\hline
    \end{tabular}
    \caption{The calculated and fitted CEF parameters from Ref.\cite{JoYb2015}. The calculated values are obtained from point-charge model while the fitted values are from fitting inelastic neutron scattering data. As can be seen the difference between the fitted and calculated B$_6^0$ and B$_6^3$ is off by a magnitude.}
    \label{Ybcefparameter}
\end{table*}

Table \ref{Ybcefparameter} lists previous works on the CEF analysis for \yto. Bertin \textit{et al.}'s results \cite{Bertin_2012} are obtained within the point-charge approximation and taken as starting parameters for the work by Gaudet \textit{etl al} \cite{JoYb2015}. They deployed a typical mean-square-root deviation minimization algorithm that searched through the six dimensional parameter space in the vicinity of the starting values from Ref. \cite{Bertin_2012}. By using a point-charge model to generate starting parameters, one assumes these starting values are close to the global minimum. However, Table \ref{Ybcefparameter} shows that the refined value from point-charge model can be different by an order of magnitude (i.e, B$^0_6$, B$^3_6$). 

\begin{figure*}[tbh]
    \centering
    \hspace*{-0.2in}
    \includegraphics[height=6in]{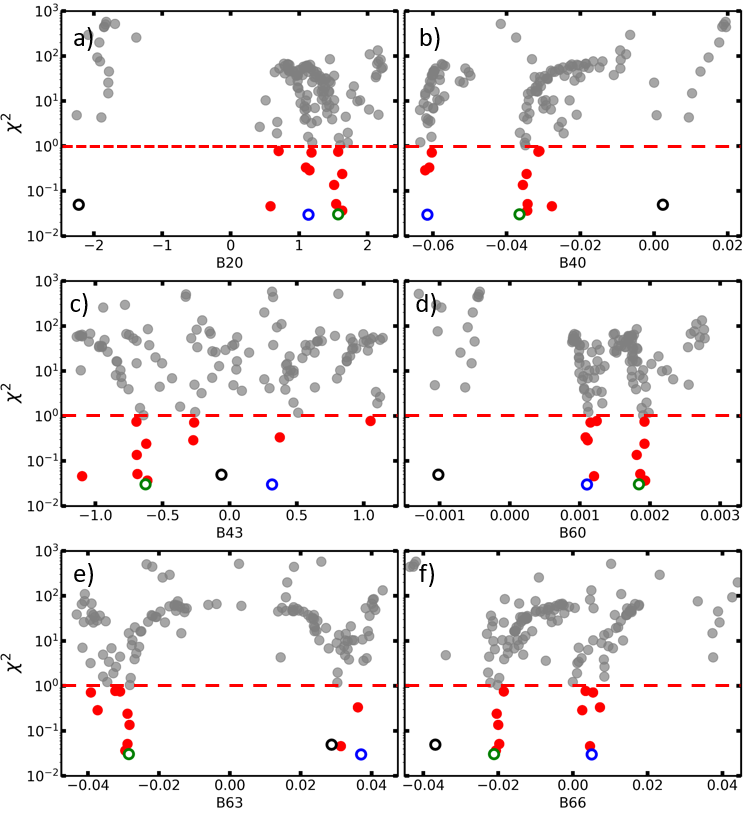}
    \caption{The fitted crystal field parameters vs $\chi^2$ for Yb$_2$Ti$_2$O$_7$. Each red point is one converged solution from the PSO with $\chi^2 <1$. The open blue circles are the 'true' solution from Ref. \cite{JoYb2015}. The open green and black circles are the second and third best solutions found by CrysFieldExplorer. In panels a), b) d) and f) we observe two separate regions in the phase space that produce acceptable results. The region where the green data points reside has not been previously reported in Ref.\cite{JoYb2015,Bertin_2012}.}
    \label{parameteryb}
\end{figure*}

Now we will demonstrate the results of CrysFieldExplorer in the \yto system. We adapt the searching algorithm using particle swarm optimization with our customized Spectrum - Characteristic Loss function discussed in previous sections. We construct the PSO model with a particle size N=400 and iteration number of 100 times. The details of the hyper-parameters can be found in Ref. \cite{PSO}. The total loss is constructed as $L_{total}=L_{\text{Spectrum}}+L{1/\chi}$, where $\chi$ in this case represents the susceptibility data. These are the same experimental data used to obtain results in Ref.\cite{JoYb2015}. We present CrysFieldExplorer's results in Figure \ref{parameteryb} and compare it with those reported in Ref.\cite{JoYb2015}. 
 
Figure \ref{parameteryb} lists all 150 results from CrysFieldExplorer on Yb$_2$Ti$_2$O$_7$ in a log scale along the y-axis. Each \yto CEF parameter is plotted against a custom defined goodness of fit $\chi^2=\dfrac{[Expected\ Value-Observed\ Value]^2}{Expected\ Value^2}$. Readers should be aware of the definition of present $\chi^2$ is simply the sum of percentage difference between true solution and calculated solution. It is calculated by comparing the excited energy levels and relative intensities, neutron spectroscopy, susceptibility and magnetization using CEF parameters provided in the Ref. \cite{JoYb2015}. This $\chi^2$ is merely an indication of the agreement goodness between calculated and true observables. We chose $\chi^2 < 1$ as a cut off number in present study as we find results below this number shows good overall agreement between fit and true solution. The solid grey circles represent 150 solutions generated using PSO algorithm. The solid red circles represent 13 solutions with $\chi^2 < 1$. The blue circles mark the parameters reported from Ref. \cite{JoYb2015} and is taken as the "true solution" in this example. The $\chi^2$ for the true solution is given an artifically 0 but shifted by 0.03 so it is visible on the log scale. One of the important observations from Figure \ref{parameteryb} is that there exists three local minima in the parameter space of \yto  crystal field. Observing Panel a), b), d), f) from Figure \ref{parameteryb} shows two out of the three local minima are concentrated regions where CrysFieldExplorer can easily converge. There is a third region that contains one acceptable solution with $\chi^2 <1$. The true solution reported from Ref. \cite{JoYb2015} marked by blue circle falls in 1 of the minima. The green and black circles are in two other minima. Interestingly, while no constraints were set on B$_4^3$ and B$_6^3$, panels c) and e) show symmetric behavior w.r.t to 0 as expected by the symmetry of local crystalline environment.

\begin{figure*}
    \centering
    \hspace*{-0.5in}
    \includegraphics[width=6.5in]{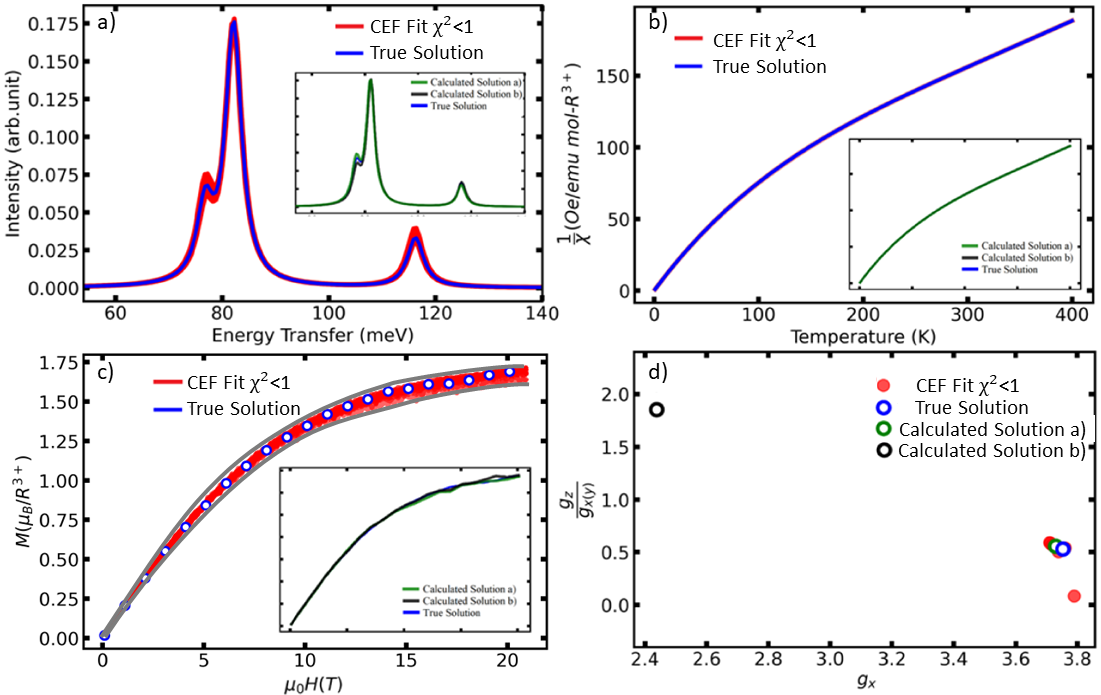}
    \caption{The bulk properties of the solutions with $\chi^2 < 1$ obtained using PSO. a) The energy vs intensity spectrum. A fluctuation $\sim$ 5$\%$ of intensity is observed around 76 meV. b) The inverse susceptibility. The calculated results almost completely overlap with the 'true' result. c)The magnetization data. Open blue circle indicates the true solution, solid red lines indicate the CEF fit. The two solid grey lines shows an region where all the solutions reside in. d) The ratio of the anisotropy g-tensors $g_z/g_{x(y)}$. The green and black data in the insets of panels a)-c) represent the solution a) and b) which are also listed in Table \ref{Ybcefparameter}.}
    \label{Ybproperties}
\end{figure*}

To analyze the physical meaning of the CEF parameters found by CrysFieldExplorer, the main panels in Fig. \ref{Ybproperties} a)-d) show the comparison between calculated and true results in terms of inelastic neutron scattering data, inverse susceptibility, magnetization and the ratio of g$_z$/g$_{x(y)}$ for the Yb$^{3+}$. Unfortunately the experimentally measured susceptibility and magnetization data was not given in Ref. \cite{JoYb2015}. We assume the calculated susceptibility and magnetization using the CEF parameters provided from Ref. \cite{JoYb2015} as "true solution".  Each red lines in a)-c) and the red data points in d) indicate the calculated properties. The blue lines and open blue circles in a) - d) represents the true solutions. The black, green color scheme indicate two other minima results from present study corresponding to the same green and black data points in Figure \ref{parameteryb}. We also list our representative results (green and black data) as CEF parameters in Table \ref{Ybcefparameter}. They are marked by a and b respectively. Clearly, Figure \ref{parameteryb} and \ref{Ybproperties} demonstrate our PSO optimizer can successfully converge to the true CEF parameters region from Ref.\cite{JoYb2015}. A universal starting range of [-100,100] meV were set for all six B$^m_n$s in Eq.\ref{hamiltonian}. For general users, we feel this range can be appropriately determined based on the magnitudes of excited energy levels. Additionally, as Spectrum - Characteristic Loss is smooth and can effectively avoid producing steep energy barriers seen in Fig. \ref{Loss}, a larger range for the CEF parameters should not be problematic. The average time for the PSO to converge to one solution in the 6-dimensional CEF parameter space was approximately 4.7 minutes on a 12th Gen Intel(R) Core i7-1265U laptop with single core processing. The maximum magnitude of the searching range compared to the order of the minimum CEF parameter B$^0_6$ is on the order of 10$^4$. CrysFieldExplorer can locate solutions within this vast range of parameter space within minutes. This is an example to show the powerful optimization capability with CrysFieldExplorer.

Observing each panel in Figure \ref{Ybproperties} reveals more details of CrysFieldExplorer's results. The main panels of a), b) and c) in Fig. \ref{Ybproperties} compare the calculated solutions of $\chi^2 < 1$ with the true solution. These solutions are over-plotted in red and form a region which makes it easier to identify the relative positions of the calculated solutions in from the true solutions. The insets of a), b) c) in Figure \ref{Ybproperties} compare the true solution with the solutions a) and b) found and listed in Table \ref{Ybcefparameter}. Solution a) and b) are indicated as green and black open circle in panel d). Readers should pay particular attention to solution a) and b) as they are in two different regions in the CEF parameter space. In panel a) of Fig.\ref{Ybproperties}, the calculated spectrum shows good agreement with the true spectrum except at $\sim$ 76 meV. We estimate approximately $\pm$ 3$\%$ variation on the peak intensities from our calculated solutions compared to the true solution. The excitations at 81.8 meV and 116.2 meV matches well in both peak width and integrated intensity. In panel b), the inverse of susceptibility shows excellent agreement as well. The CEF fit and true solution are indistinguishable by eye. Panel c) shows the 10K magnetization data. The solid grey lines mark an region where the true solution and calculated solutions with $\chi^2 < 1$ reside in. The inset of panel a), b), c) shows a similar story, no significant difference between true solution and solutions a) and b). Panel d) is the most interesting plot as it shows the ratio of the anisotropy g-tensors $\dfrac{g_z}{g_{x(y)}}$. Only in this panel we can distinguish solution b) from solution a) and the true solution. Observing panel d), most of the calculated anisotropy g-tensors are in one dominant region where the ratio of $g_z/g_{x(y)}$ is $\sim$ 0.5 while $g_x$ is close to 3.8.  Our best solution, represented by the green open circle, yields a set of g-tensor $g_{x(y)}$ = 3.73 $g_{z}$ = 2.07 that are within the uncertainty given from the published result of 3.69 $\pm$ 0.15 and 1.92 $\pm$ 0.20 \cite{JoYb2015}.The ratio of $g_z/g_{x(y)}$ is of particular interest. Previous polarized neutron studies from Cao \etal \cite{Cao_2009} reported $g_x(y)=4.1$ and $g_z=2.25$ resulting in a ratio of 1.82. Bertin's \cite{Bertin_2012} and Gaudet's\cite{JoYb2015} results are listed in Table \ref{Ybcefparameter} with $g_z/g_{x(y)}$= 2.0 and 1.92. From our solution a), the ratio can be calculated to be 1.79 closes to Cao \etal's result. Now we turn our attention to solution b). Despite having a good agreement with the experimental values in the bulk property measurements, it shows a strong ratio of $g_z/g_{x(y)}$ $\sim$ 3.5, with $g_z=4.51$ significantly larger than solution a) and the true solution. This observation suggests solution b) represents a local minimum that fits well with limited constraints. By analyzing the anisotropy g-tensor, we conclude that solution b) does not agree with the physical model of \yto.

Now we turn our focus to solution a) which is in a different region with the true solution, but we cannot distinguish solution a) from the true solution by all the given measurements. Table \ref{Ybgroundstate} compares the ground state wavefunctions between solution a) marked by green circle in Figure \ref{parameteryb} and that reported in Ref. \cite{JoYb2015}. Both solutions also produce similar ground state wavefunctions as reflected by the similarity in the anisotropy g-tensors. The dominant term in the ground state wave function from both solutions are the $\pm\dfrac{1}{2}$ terms. 

Combining these results with previous discussions suggests solution a) is a degenerate solution under the existing constraints imposed by the experimental data reported in the \yto system. Only one set of CEF parameters were reported in each paper from previous studies \cite{huiboPRL,Bertin_2012,JoYb2015}, but our discovery implies that these papers may need more data to conclusively determine the CEF parameters. The ground state of \yto has been extensively studied from various perspectives and our results do not imply whether the previous analyses are incorrect. However, it is an alarming discovery that suggests future CEF analysis should be cautious in confirming or denying their findings. In next chapter, we will show that lack of enough constraints may pose serious uncertainty in determining the true CEF parameters. 

\setlength{\tabcolsep}{10pt}
\def\arraystretch{2.2}
\begin{table*}[]
    \centering
    \begin{tabular}{c c c c c c c c c } 
    \hline\hline
        $J_z$  &  $\dfrac{-7}{2}$ & $\dfrac{-5}{2}$ & $\dfrac{-3}{2}$  & $\dfrac{-1}{2}$ & $\dfrac{1}{2}$ & $\dfrac{3}{2}$ & $\dfrac{5}{2}$ & $\dfrac{7}{2}$ \\
    \hline
         GS eigenfunctions\cite{JoYb2015}& 0 & 0.0866 & 0 & 0 & $-0.9283$ & 0 & 0 & 0.3616 \\
         
                                 & $-0.3616$ & 0 & 0 & $-0.9283$ & 0 & 0 & $-0.0866$ & 0 \\
         \hline
         GS eigenfunctions\footnote{The parameters indicated by green circles in Figure \ref{Ybcefparameter}} & 0 & $-0.076$ & 0 & 0 & $-0.924$ & 0 & 0 & $-0.375$ \\
          & 0.375 & 0 & 0 &$-0.924$ & 0 & 0 & 0.076 & 0 \\

    \hline\hline
    \end{tabular}
    \caption{The ground state eigenfunctions from ref.\cite{JoYb2015} and present work of solution a) as reported in Table \ref{Ybcefparameter}. Despite the difference in the CEF parameters, the ground state eigenfunctions looks similar with dominant $\pm\dfrac{1}{2}$ terms.}
    \label{Ybgroundstate}
\end{table*}

Regardless of which set of CEF parameters is correct, one of CrysFieldExplorer's biggest advantages is the capability to directly commence the fitting procedure starting from Eq. \ref{Ybhamiltonian} without any \textit{a priori} knowledge of the starting parameters. This allows users to skip the step of building a point-charge model or conduct Monte Carlo simulation, either of which can be daunting for inexperienced users. CrysFieldExplorer only requires a wide fitting range to be given. All the parameters of \yto were chosen from a range of [-100, 100] meV which can be estimated from the magnitude of the leading B$_2^2$ term. This range requirement can be easily estimated by users. Too large or too small beyond this range could result in the excited energy levels to be unreasonable large or small. 

To conclude, in this section we demonstrate that CrysFieldExplorer can quickly produce a list of CEF solutions for \yto including the ones reported in the literature. Our improved loss function $L_{\text{Spectrum}}$ has a higher sensitivity to the change of CEF parameters and a better global structure that allows us to search through a wide range of parameter space. Identifying which one is the real solution is more complicated. On the one hand, there may not be enough constraints to conclude which solution is the correct one. On the other hand, as most of the experimental data are collected from powder sample, the information about the relationship between local coordinates and crystal structure is intrinsically missing, thus rendering the the model unable to reduce further to one result.

\subsection{Analysis on the Crystal Field Parameters of the Low Symmetry Tripod Kagome Lattice $Er_3Mg_2Sb_3O_{14}$}

\begin{figure*}
    \centering
    \includegraphics[height=5in]{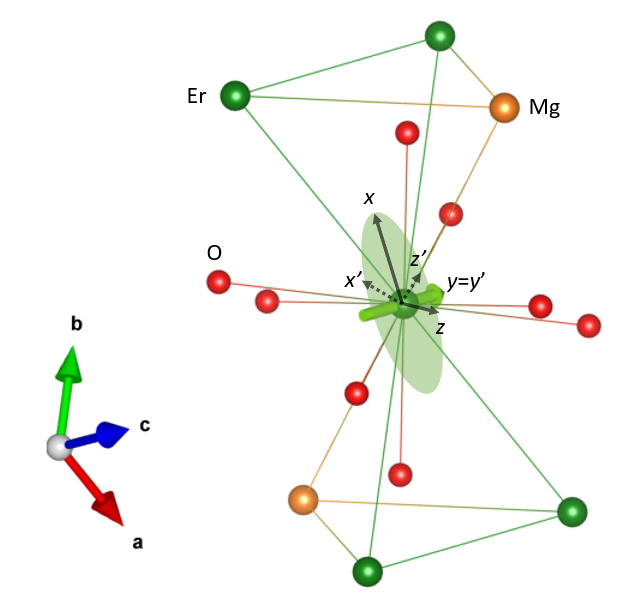}
    \caption{The local structure of \EMSO. Green atoms indicate Er, yellow atoms are Mg and the red atoms are Oxygen. The green arrow on the center Er atom indicate the $C_2$ rotational axis and is chosen as the y-axis. The green surface is perpendicular to the y-axis. The direction of x and z axis can be arbitrarily chosen within this plane while preserving the form of CEF Hamiltonian in Eq. \ref{HamiltonianEr}.}
    \label{ErStructure}
\end{figure*}

Now we consider a more challenging case - a tripod Kagome lattice that hosts various exotic magnetic ground states depending on the constituent rare earth element \cite{zhilin}. We picked $Er_3Mg_2Sb_3O_{14}$ as the test case here. The CEF fitting becomes more challenging for a magnetic ion at low point-symmetry site because the number of none zero crystal field parameters increases. Fig. \ref{ErStructure} shows the local structure of an Er atom in the tripod Kagome lattice Er$_3$Mg$_2$Sb$_3$O$_{14}$ with a space group of $R\Bar{3}$m. The rare earth ion Er$^{3+}$ is surrounded by eight oxygen atoms similar to that in pyrochlore system. However, the local site symmetry becomes $C_{2h}$ in the tripod Kagome lattice, much lower than $D_{3d}$ in the pyrochlore lattice. The twofold rotational $C_2$ axis lies in the Kagome plane and is displayed in Fig. \ref{ErStructure} as the green arrow on the center Er atom. As a result, this system requires 15 CEF parameters to fully describe the crystal field Hamiltonian,
\begin{equation*}\label{HamiltonianEr}
\begin{aligned}
H_{CEF}= &B^0_2\hat{O}^0_2 + B^1_2\hat{O}^1_2 + B^2_2\hat{O}^2_2 + B^0_4\hat{O}^0_4 + B^1_4\hat{O}^1_4 + B^2_4\hat{O}^2_4\\
& + B^3_4\hat{O}^3_4 + B^4_4\hat{O}^4_4 + B^0_6\hat{O}^0_6 + B^1_6\hat{O}^1_6 + B^2_6\hat{O}^2_6 + B^3_6\hat{O}^3_6 \\
&+ B^4_6\hat{O}^4_6 + B^5_6\hat{O}^5_6 +B^6_6\hat{O}^6_6. 
\end{aligned}
\end{equation*}

According to Hund's rule, the total, spin and orbit angular momentum for Er$^{3+}$ are $J$=15/2, $S$=3/2 and $L$=6 respectively. This results in 2$J$+1=16 fold degeneracy for the ground state of Er$^{3+}$. Since Er$^{3+}$ is a Kramer ion, all 16 levels are doublets, 7 excited states are expected to be observed from inelastic neutron scattering measurement. Dun \textit{et al.} \cite{zhilin} observed 6 out of 7 CEF transitions at 6.4(2), 10.5(3), 21.6(4), 50(1), 65(1) and 67.5(9) meV. Due to the large amount of none zero CEF parameters, they were not able to directly fit the CEF transitions with currently existing software packages such as PyCrystalField. A list of CEF parameters was obtained from an effective point-charge model which effectively reduced the number of fitted parameters. 

Now we demonstrate how CrysFieldExplorer can help provide insight to this problem and discuss why directly fitting the CEF parameters with given experimental data was not feasible. Due to the low site symmetry in Er$_3$Mg$_2$Sb$_3$O$_{14}$, the parameter space is magnitudes larger than that in \yto and traditional mean-square-root loss function is less sensitive to the 15 CEF parameters comparing to $L_{\text{Spectrum}}$ as demonstrated previously in Fig. \ref{Loss}. To handle this case, we need to improve the optimization from our previous particle swarm optimizer. We continue apply the $L_{\text{Spectrum}}$ and deploy the Covariance Matrix Adaptation Evolution Strategy (CMA-ES) \cite{hansen2006cma}, which converges rapidly on complex optimization problems where the global minimum is extremely sharp in the parameter phase space. 

For a local site with low point group symmetry such as monoclinic in the present case, the choice of principle $xyz$ axis can be somewhat arbitrary. In the case of tripod Kagome lattice, the $y$-axis is chosen to be parallel to the two-fold rotational axis $C_2$ such that none zero terms of the CEF parameters are expressed in Eq. \ref{hamiltonian}. The choice of x and z axis is arbitrarily defined within a perpendicular plane to the $C_2$ rotational axis. The x-z plane is indicated as the green plane in Fig. \ref{ErStructure}. The dashed and solid arrows indicate x and z axis are not fixed within the plane perpendicular to y-axis.  Rudowicz \etal \cite{rudowicz1986standardization, rudowicz1985relations} has demonstrated that within the 15 CEF parameters, certain rotations exists along $y$-axis that make one of the CEF parameters 0 . Here we deploy the transformation of rotating the CEF coordination system w.r.t local y-axis by an angle $\alpha$ such that $B_2^0$ is 0 for all the solutions. The rotational angle $\alpha$ is defined as,
\begin{equation}
    tan(2\alpha)=\dfrac{B_2^1}{3 \times B_2^0-B_2^2}.
    \label{alpha}
\end{equation}

The rotational transformations of relevant CEF parameters have been tabulated in Table 2 from Ref.\cite{rudowicz1985relations}. The CEF parameters reported here have all been transformed under this standardization rule with each set of parameters rotated by an angle $\alpha$. A complete list of CEF parameters and the transformation angle $\alpha$ are plotted in SFig. 1 in the supplemental material. 

\begin{figure*}
    \centering
    \includegraphics[width=6.5in]{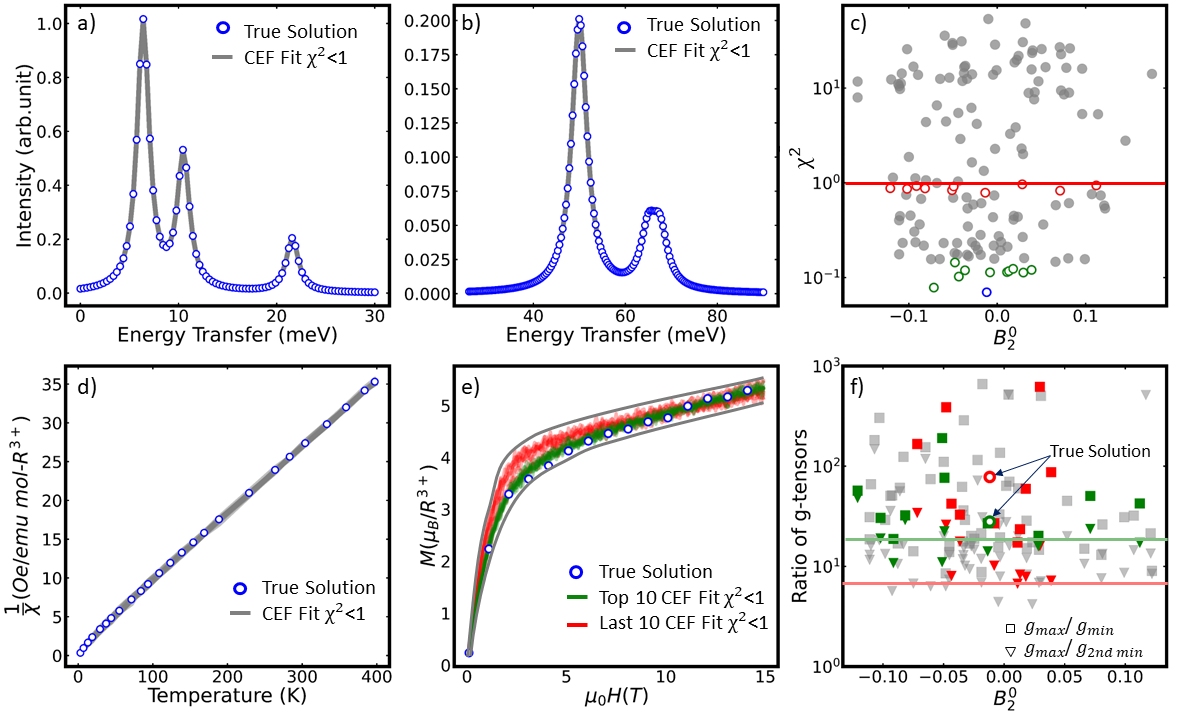}
    \caption{a), b) Neutron spectroscopy data plotted within the energy transfer range of [-1, 33] meV and [25,90] meV. c) Solid grey points are the complete solutions. The red horizontal line marks $\chi^2=1$. Open red circles are the last 10 solutions with $\chi^2<1$. Open green circles are the top 10 solutions. Open blue circle is the true solution. The y-axis is plotted in log style. d) Inverse susceptibility of the true solution and solutions with $\chi^2 < 1$. e) Magnetization of the true solution, top 10 solutions and last 10 solutions with $\chi^2 < 1$. The solid grey lines are estimates of the upper and lower bounds of the CEF fit with $\chi^2 < 1$. f) Ratios of the anisotropy g-tensors. Squares indicate $g_{max}/g_{min}$, triangles indicate $g_{max}/g_{2nd\ min}$. Grey data points indicate all solutions with $\chi^2 < 1$. Green and red data points show the top and last 10 CEF fits corresponding to panel c). All the plotted solutions produce Ising-like anisotropy g-tensors.}
    \label{Erproperties}
\end{figure*}

We summarize our results in Fig. \ref{Erproperties}. Different to the \yto case, CrysFieldExplorer can find a large number of solutions across the parameter space with no obvious segregation. We first look at Fig. \ref{Erproperties} c). As \EMSO has 15 (14 independent) CEF parameters, we choose the $B_2^0$ as an example and plot the custom defined $\chi^2$ loss on a log scale. The solid grey circles indicate all 150 solutions produced by CrysFieldExplorer in a time span of approximately 12 hours. We chose 1 as a cutoff number for the custom defined $\chi^2$ and denote the best and worst 10 solutions with $\chi^2 <1$ using green and blue open circles. Meanwhile, open blue circles indicate the true solution. The $\chi^2$ for the true solution is artificially defined as 0 but shifted by a small number of 0.03 in order to be shown on the log plot.

Figure \ref{Erproperties} a) and b) show the neutron data on the energy transfer range from -1 to 33 meV and 25 to 90 meV respectively. Fig. \ref{Erproperties} d) is the inverse susceptibility data. The solid grey lines in these three plots are an overlay of all the solutions with $\chi^2 < 1$ found by CrysFieldExplorer using CMA-ES algorithm optimizing $L_{\text{Spectrum}}$. The open blue circle indicates the solution taken from Ref.\cite{zhilin} which are treated as true solution. Fig. \ref{Erproperties} a) b) and d) are the three data sets used by CrysFieldExplorer during the optimization process.

Figure \ref{Erproperties} e) compares the calculated magnetizations with the true solution. The solid red lines and solid red lines mark the top and last 10 solutions with $\chi^2 < 1$. Following the same color scheme, open blue circles are the true solution. We drew two solid grey lines to indicate a region where all the solutions stay within. The determination of the grey lines are guided by eye. 

Figure \ref{Erproperties} f) shows the ratio of the anistropy g-tensors. The grey squares indicate the ratio between the largest and smallest while the pink triangle indicate the largest and 2nd smallest anisotropy g-tensors for all solutions with $\chi^2 < 1$. The green and red square represent the top and last 10 CEF fits with $\chi^2 < 1$.

Figure \ref{Erproperties} reveals several important features of CrysFieldExplorer when applied to the \EMSO system. CrysFieldExplorer can perform perfect fits to the neutron data in panels a) and b) to the degree that is impossible for the human eyes to distinguish. We performed fitting on both the discrete CEF energy levels and their relative intensity as well as fitting directly on the neutron spectrum by applying a resolution limited Lorentzian function to compute the entire spectrum. There are 73 out of 150 sets of parameters with $\chi^2 < 1$, all of them produced perfect match to human eyes between the true solution and CEF fit. Additional constraints were imposed when adding the inverse susceptibility data to the fit. However, within the $\chi^2 < 1$ range, CrysFieldExplorer was still able to discover multiple solutions that fit all given experimental observables well. 

The powder averaged magnetization data was not considered during the fit process but used as a benchmark to check the fitting results. Comparing magnetization data in panel e) of Fig. \ref{Erproperties}, we start noticing differences between each solution set. The top 10 solutions show good match to the true solution at all magnetic fields measured. However, the last 10 CEF fitted solutions with $\chi^2 < 1$ begin to show small deviations at 4 T magnetic field. However, we argue our these 10 solutions still fit all the data well and could still be considered acceptable a few decades ago given the computing power at that time. This shows the efficiency and accuracy CrysFieldExplorer exhibits when being deployed on the latest high-performance computers.

Figure \ref{Erproperties} f) is particularly interesting because it indicates all of the solutions produced by CrysFieldExplorer are Ising-like, that is one of the three $g_x$, $g_y$ or $g_z$ are dominant comparing to the other two. We do not distinguish $g_x$, $g_y$ or $g_z$ here because there is a certain degree of freedom in choosing the local frame. From present data, it is not possible to obtain the relationship between local frame and global frame. Similar to the \yto case, the powder data measured with the mentioned techniques do not have crystal structure resolution. 

To conclude, our results show that with increased free CEF parameters, the constraints required to narrow down to a single solution also increases drastically. CrysFieldExplorer was able to produce 73 solutions out of 150 searches that matches neutron data and susceptibility data well. The powder averaged magnetization data were not used in the fitting because it is computationally expensive to calculate the powder average. However, single crystal magnetization data could be added in the fitting without much penalty on the computational cost to impose more constraints. Despite not being able narrow down to one or a few solutions, CrysFieldExplorer is able to produce a list of acceptable solutions. These solutions produce excellent agreement between the observed and measured data such as inelastic neutron scattering spectrum, susceptibility, and magnetization measurements. We demonstrate in the case of \EMSO, neutron spectroscopy and susceptibility do not impose enough constraints to fully describe the problem, thus making it an under-determined case when only considering limited experimental observables. Therefore, we agree with the statement in the original paper that direct fitting of the CEF parameter is not possible \cite{zhilin}. Nonetheless, in the \EMSO system, by wrapping $L_{\text{Spectrum}}$ into CMA-ES, we show the strong capability and high efficiency CrysFieldExplorer exhibits. The approximated time span to generate 150 sets of solutions is approximately 12 hours. We argue 73 of the 150 solutions produce acceptable results. Future updates can further improve the efficiency by leveraging high performance computing on multi-core processors on user-end computers. CrysFieldExplorer's results also suggest that when fitting a large number of CEF parameters, user should exercise extra caution when choosing the correct solution. In some of the previous research, a solution is believed to be valid when the neutron scattering data, susceptibility data, magnetization measurements, anisotropy g-tensors, etc are all consistent. However, utilizing CrysFieldExplorer's efficient algorithm, we show that these condition can all be met with very good agreement between experimental and fitted results in low site symmetry materials such as \EMSO with 15 (14 independent) none zero CEF parameters. As the number of CEF parameter increases, more experimental data should be acquired to fit the CEF Hamiltonian with experimental model to obtain conclusive results. These can be, but not limited to, polarized neutron diffraction, specific heat etc. Future researchers should be cautious to draw conclusions on CEF analysis when considering limited experimental data.

\section{Implementation}

We aim to provide the most straight forward implementations for CrysFieldExplorer to the research community. Users first need to specify the magnetic ions and local site symmetry. This allows CrysFieldExplorer to generate a list of none-zero CEF parameters. Then the total, orbital and spin angular momentum $J$, $L$ and $S$ can be automatically determined. The program will construct the Hamiltonian up to 6th order using the Stevens operators as well as the matrix representation of $J_{plus}$, $J_{minus}$ and $J_z$. Then users can import the experimentally measured data into the program for fitting. In the present version of CrysFieldExplorer, it is able to fit neutron scattering data, susceptibility, single crystal magnetization, anisotropy g-tensor data all together with different weights. For neutron scattering data, relative intensities of the observed transitions from the ground state need to be specified by users from a data reduction software. The currently required data format is a two-column text file. The experimental data will be loaded into the optimization program. For high-symmetry cases the program will select particle swarm optimization and for low-symmetry system a more efficient CMA-ES will be chosen instead. The loss function will be predetermined as discussed before. Although CrysFieldExplorer does not require a set of specific starting parameters, a suitable range for all CEF parameters still needs to be given. However, this can be estimated from the magnitudes of the transition energy levels from the inelastic neutron scattering. The optimization strategy will minimize Spectrum - Characteristic Loss first. Once Spectrum - Characteristic Loss reaches -10, it will start optimizing magnetic measurements or any other user defined data. Users are able to prioritize different data by fine tuning their respective weights.  By default, CrysFieldExplorer will generate 100 converged CEF parameters in .csv file. These parameters can be checked with either CrysFieldExplorer or a 3rd party software for consistency. The detailed implementation of CrysFieldExplorer can be found at https://github.com/KyleQianliMa/CrisFieldExplorer.

\section{Use and Limitations}
The program CrysFieldExplorer presented here provides a fast converging, gradient free method that scans through large CEF parameter phase space to provide satisfactory results based on user input. It bypasses the need of determining initial parameters from point-charge model which can be difficult for inexperienced users. CrysFieldExplorer contains two optimization methods, PSO and CMA-ES, to minimize a set of specially designed loss functions. The choice of each optimization method is generally chosen such that for CEF parameters $<6$, PSO is chosen, otherwise CMA-ES is preferred. Although CrysFieldExplorer can successfully produce reliable solutions, it cannot interpret the physical meaning of each solution. Furthermore, if the problem is under-defined, that is experimental data does not pose enough constraints to the CEF Hamiltonian, multiple solutions produced by CrysFieldExplorer can be confusing to determine which solution is the correct solution. Although we provide two optimization methods, there has not been a standard way to choose the best optimization methods for different problems. The weight for different losses can also be tricky to balance. In some test cases it is easier to fit one experiment than another. Finally, the current version of CrysFieldExplorer does not provide a user interface. It requires users with basic knowledge of Python programming language.

\section{Conclusion}
We have developed a lite python-based program CrysFieldExplorer to provide an efficient procedure of fitting the CEF parameters from experimental data. CrysFieldExplorer uses two evolutionary algorithms - particle swarm optimization (PSO) and Covariance Matrix Adaptation Evolution Strategy (CMA-ES) to minimize a newly designed loss function - Spectrum Characteristic loss $L_{\text{Spectrum}}$. Comparing $L_{\text{Spectrum}}$ function with traditional mean-square-root loss, we showed that $L_{\text{Spectrum}}$ is more sensitive to input parameters, significantly reduces the steepness of energy barriers on the parameter-loss space and has a better global structure around local and global minima. This allows CrysFieldExplorer to start the fitting procedure directly from the CEF Hamiltonian without prior knowledge of starting parameters. Although a fitting range of the CEF parameter is required, the range can be generously given such that it matches the magnitudes of the excited CEF energy levels. To demonstrate the use of CrysFieldExplorer, we have provided two examples in two different systems to show that our algorithm can discover multiple solutions that matches experimental data equally well. Our findings suggest that as the number of to-be-fitted CEF parameters increases, traditional experimental observables such as neutron scattering data, susceptibility and magnetization measurement may not impose enough constraints to fully describe the problem. Furthermore, neither the powder neutron spectroscopy nor the powder susceptibility data have crystal structure resolution to define the connection between local ion and crystal structure frame. This arbitrariness could further produce confusion when determining the correct CEF fit. Therefore, the present standard of reporting CEF parameters may not be conclusive as the CEF parameter phase space is highly degenerate. When conducting CEF analysis with limited experimental data, further measurements such as polarized neutron diffraction can provide additional information to help better determine the correct CEF Hamiltonian. 

\section{Acknowledgments}

The research at Oak Ridge National Laboratory (ORNL) was supported by the U.S. Department of Energy (DOE), Office of Science, Office of Advanced Scientific Computing Research under the contract ERKJ387, and Office of Basic Energy Sciences, Early Career Research Program Award KC0402020, under Contract DE-AC05-00OR22725. A portion of this research used resources at the High Flux Isotope Reactor and the Spallation Neutron Source, DOE Office of Science User Facilities operated by the Oak Ridge National Laboratory.

%

\end{document}